\documentstyle[12pt]{article}
\newcommand{\be}{\begin{equation}} \newcommand{\ee}{\end{equation}} 
\newcommand{\bea}{\begin{eqnarray}}\newcommand{\eea}{\end{eqnarray}}

\begin{document}
\begin{titlepage}
\begin{flushright}
IP/BBSR/96-71 \\
hep-ph/9612418 \\
\end{flushright}
\title {Hybrids as a Signature of Quark-Gluon Plasma} 
\vskip 3.0 cm 
\author{ {\bf Afsar Abbas}$^{1}$ and {\bf Lina Paria}$^{2}$ \\
Institute of Physics, Sachivalaya Marg, \\
Bhubaneswar-751005, India.} 
\footnotetext[1]{e-mail: \ afsar@iopb.ernet.in} 
\footnotetext[2]{e-mail: \ lina@iopb.ernet.in}
\maketitle
\thispagestyle{empty}
\begin{abstract}
We show that the dynamics of the Quark-Gluon Plasma is such that during
hadronization the creation of hybrids will predominate over
the creation of mesons, giving
a novel signature of the existence of QGP. 
At $ T = 0 $ the $(q\bar{q}g)$ hybrids are known to decay 
strongly into a pair
of mesons. We find that at temperatures relevant 
to the QGP, this channel is forbidden. 
This would lead to significant
modifications of the photonic signals
of the QGP.
\end{abstract}
\end{titlepage}
\eject

\eject

\newpage

In addition to the conventional mesons and baryons, on very general
arguments based on QCD one expects non-conventional systems like 
glueballs (made up of two or more gluons) and hybrids (like $ q\bar{q}g $).
There have been some claims that these glueballs and hybrids may 
already have been seen in the laboratory [1-4]. Here we concentrate upon
the ($ q\bar{q}g $) hybrid [1,2]. These are expected to decay strongly into
a pair of mesons [1,2]. More work has been done recently [3,4] to
understand the structure of these objects and their possible 
experimental identification in low energy hadron spectroscopy.
The hybrids remain an exciting and open problem.

Is there any other place where hybrid may manifest themselves more
explicitly? Perhaps finite temperatures may hold the clue? With
this in mind we look at the Quark-Gluon Plasma (QGP) as a possible
laboratory for studying the hybrids. The field of QGP is currently
a very active area of research, both experimentally as well as
theoretically (see ref. 5 \& 6 for recent reviews). It is extremely
important to be able to identify QGP if and when it is
formed. Several signature of QGP have been suggested [5,6] and looked
for, but none of the suggested signals have been unambiguous to
demonstrate in a clearcut manner whether QGP has been formed in
the laboratory or not. Hence we ask the question- could a hybrid
be that signal of QGP? We shall show that indeed it may very well be so!

In this paper we consider a system of quarks, antiquarks and gluons
placed in a heat bath with which it can exchange the energy and
particle number. We study the thermodynamics of the system confined in 
a MIT bag [7]. All the thermodynamic quantities are obtained by
using discrete single particle states of the quarks and gluons which
are obtained by solving the equation of motion with linearised boundary
conditions [8].

The number of quarks and gluons of the system at finite temperature
$ T $ is given by

\be
N_{q} \ = \ d_{q} \ \sum_{i} { g_i} {f_i}
\ee
\noindent

\be
N_{g} \ = \ d_{g} \ \sum_{i} { g_i} {b_i}
\ee
\noindent

where $ f_i \ = \ {1 \Big{/}{\Big(} \ e^{(\epsilon_{i}- \mu )/T} 
\ + \ 1{\Big)}} $ , the F - D distribution function for quarks and

$ b_i \ = \ {1 \Big{/}{\Big(} \ e^{\epsilon_{i} /T} 
\ - \ 1{\Big)}} $ , the B - E distribution function for gluons, 
$d_{q} = 2.3/2 $ and $ d_{g} = 8 $. Here in the case of the quarks, 
the factor of 2 is for the
flavours, 3 for the color and the division by 2 is for 
the particular spin of
q and $ \bar{q}$. For the gluon case, 8 comes from the 
color configuration and
$ g_{i} $ is the spin degenaracy factor for the $ i^{th} $ single particle
state with energy  $ \epsilon_{i} $ and $ \mu $ being the chemical potential
for quark.  All the quantities can be calculated for the antiquark
by replacing $ \mu $ by $ - \mu $. As our aim is to study
the thermodynamic properties of the hybrid ($ q\bar{q}g $), we ensure
$ N_{q} = 1$ and $ N_{\bar{q}} = 1 $. The gluon arises due to the thermal
excitations as the temperature $ T $ increases.

The energy and the free energy of the quark and gluon is given by

\be
E_{q} \ = \ d \ \sum_{i} { g_i} {\epsilon_{i}} {f_i}
\ee
\noindent

\be
E_{g} \ = \ 8 \ \sum_{i} { g_i} {\epsilon_{i}} {b_i}
\ee
\noindent

\be
F_{q} \  =  - \ d \ T \ {\sum_i}  \ {g_i}  
\ ln {\Big (} \ 1  \ +  \ e^{- ({\epsilon_i - \mu ) /T}} {\Big)} 
\ee
\noindent

\be
F_{g} \  =  \ 8 T \ {\sum_i}  \ {g_i}  
\ ln {\Big (} \ 1  \ -  \ e^{- {\epsilon_i /T}} {\Big)} 
\ee
\noindent

where $d $ is the quark degeneracy factor.
Now adding the well known zero point energy term $ (BV + C/R) $ [7]
we get the total free energy of the system as

\be
F \ = \ F_{q} \ + \ F_{\bar q} \ + \ F_{g} + 2 \mu  \ + \ BV \ + \ C/R
\ee
\noindent

The pressure generated by the participant gas ( q, $ \bar{q} $, g) is

\be
P  \  =  \ - \left ({\partial F \over 
{\partial V}}\right)_T 
\ee
\noindent 

This is balanced by the bag pressure constant $B$ leading to the stability 
condition of the system. From this equillibrium condition one gets
the total energy $ E = 4BV $.

We calculate the free energy of the system of a quark, an antiquark and
gluons from equation(7) which gives a minimum at a particular value of
$R$ arising from the equillibrium condition of the system. At low
temperatures the gluonic contribution is small but at higher 
temperatures which are relevant to QGP, a single gluon arises due
to the thermal excitations giving rise to a hybrid
($ q\bar{q}g $).

Note that our picture of the hybrid at finite temperatures is
akin to what Close and Page [3] are considering i.e. it's a
gluonic excitation of meson. So what may have been meson at 
$ T = 0 $ converts into a hybrid at higher temperatures due
to the gluonic excitation arising therein. The flux tube model used in
ref.[3] had earlier been developed and used in the case of
the glueball [9] and the hybrid [10].

Next we suppress the gluonic part to get a pure
mesonic system $ (q\bar{q})$ whose energy, free energy etc. changes
with temperature giving the minimum of $F$ at a particular value
of $R$. The value of $ C $ taken for pure mesonic system is
$ \sim 0.04 $ whereas for $ (q\bar{q}g)$ system, it is $ \sim 0.4 $ [8].

We take $ B^{1/4} = 250 $ MeV, and then the temperature of the
system of q, $ \bar{q}$, and g is increased. We calculate the free
energy of the system which has a minimum at a certain value of the
bag radius $ R $ and notice the variation of it with temperature.
At a particular $ T $ at which a single gluon attaches 
to the quark, antiquark
system leading to a hybrid, we note the corresponding equillibrium radius
of the hybrid at which $F$ is minimum. This gives the corresponding
energy of the hybrid. Next we calculate the same thermodynamic
quantities at the same $T$ for the pure mesonic system. 
The free energy for the meson and the hybrid cases as a function
of radius $ R $ are displayed in Fig.1. The equillibrium radius
and the energy are given in Table 1.

Looking at Fig.1. we notice that if a mesonic bubble is created at
$ T = 183 $ MeV, then depending upon whether the radius was smaller or
larger than $ 0.658 $ fm, the bubble will grow or shrink to attain the
equillibrium radius to minimise the free energy of the system. But
now the temperature is enough to excite a thermal gluon. As we can
view our hybrid $ (q \bar q g) $ to be the gluonic excitation of the
meson [3], the gluon will sink in to make the meson into a hybrid
of radius $ 0.807 $ fm thereby reducing the free energy of the system.
If a hybrid of radius greater (or smaller) than $ 0.807 $ fm was created,
then it would shrink (or expand) to attain the equillibrium radius.
Thus one notes that the creation of hybrids would be favoured over the
creation of mesons at $ T = 183 $ MeV for $ B^{1/4} = 250 $ MeV.
Also note that  $ M_{H}\ < \ 2 m_{M} $.

Similarly for (i) $B^{1/4} = 200 $ MeV, we get $ m_{H} = 3.637 $ GeV with
the radius $ R = 1.014 $ fm at the temperature $ T = 146.5 $ MeV, whereas for
the pure mesonic case, $ m_{M} = 1.944 $ GeV with the radius $ R = 0.823 $ fm.
 (ii) for $B^{1/4} = 300 $ MeV, we get $ m_{H} = 5.313 $ GeV with the
radius $ R = 0.67 $ fm at the temperature $ T = 219.5 $ MeV, whereas for the
pure mesonic case, $ m_{M} = 2.907 $ GeV with the radius $ R = 0.548 $ fm.

It is well known that in the finite temperature case of QGP, one may
use the bag constant $ B$ as parameter [ 6,p269]. Hence there 
may be uncertainty in the quantitative prediction of the model as 
evident above. But there need be no ambiguity regarding the qualitative
prediction. Qualitatively we notice that for any reasonable value of
$ B$ at finite temperatures, the creation of the hybrids
will predominate over the creation of the mesons and also
that  $ m_{H}\ < \ 2 m_{M} $ holds.

How come at $ T = 0 $ the hybrid is above two meson production
threshold [1,2] and hence decays strongly into these channels, but
at finite temperature $ m_{H}\ < \ 2 m_{M} $ ? The answer lies in the
temperature dependence of the hybrid and the meson masses in our model
calculation. The situation is akin to the $ \sigma $ and $ \pi $
cases as studied by Hatsuda \& Kunihiro [11]. They studied the QGP
to hadron phase transition as an extension of the NJL model. They showed that
the mass of the $ \sigma $ meson ($ m_{\sigma}$) decreases and the mass 
of the pions ($ m_{\pi} $) increases with temperature. There is a
temperature $ T_{\sigma} \sim 190 $ MeV at which 
$ m_{\sigma}\ < \ 2 m_{\pi}$. Thence the decay width 
of $\sigma \rightarrow  2 \pi$
goes to zero, indicating no decay mode of $ \sigma $ into $ 2 \pi $
at that temperature. Hence at this temperature $ \sigma $ decays
electromagnetically only. Similarly in our calculation (table.1.)
at some temperature
$ T \sim 183 $ MeV, we get $ m_{H}\ < \ 2 m_{M} $ which
implies that the decay channel to two mesons is forbidden. 
So the hybrids can
decay electromagnetically producing photons.

So what does our calculation have to say about the hadronisation
of QGP ? In the standard picture of QGP consisting of equal number
of quarks and antiquarks, as the system hadronises one would
expect it to go to a system of mesons [5,6]. However our results
here suggest that the dynamics of the system is such that
the system would prefer to create
hybrids rather than mesons. In an ideal situation, only hybrids
would be created. In any condition, the bulk of the matter should
go to the hybrids. As such we suggest that the hadronisation
of the QGP through the hybrids should be taken as a signal of
the existence of QGP.

As $ m_{H}\ < \ 2 m_{M} $, the hybrid can not decay into
a pair of mesons ( as they do at $ T = 0 $ [1,2] ). Hence
they decay electromagnetically as $ H \rightarrow meson \ + \ photon $.
The basic process is 
$ q(\bar{q}) \ + \ g \rightarrow \ q(\bar{q}) \ + \ \gamma $.
On dimensional grounds, we calculate the decay width to be 
$ \Gamma_{decay} \ \sim \  {\alpha \alpha_{s}} /{m^{2} V} $.
Where $ \alpha \ = \ 1/137 $ and $ \alpha_{s} \ = \ 0.2 $.
$V $ is the volume in which the decay process takes place(which
we take as the volume of the bound state) and $ m $ is the mass of
the hybrid. With these values 
$ \Gamma_{decay} \sim 0.4 \times 10^{18} sec^{-1}$
(If we take at the $ B^{1/4} \ = \ 200 $ MeV case, we get 
$ \Gamma_{decay} \sim 0.3 \times 10^{18} sec^{-1} $). Hence on the whole
we expect the hybrid life time to be $ \sim 10^{-18}$  
 sec. This time is much larger than the hadronisation
time of the QGP. This will delay creation of mesons substantially.
In the mixed phase of QGP, the hybrids are likely to be predominant.

As we showed above, the hybrid state decay electromagnetically as
$ H \rightarrow meson \ + \ photon $ 
(e.g. $ 0^{-+} \rightarrow b_{1} \ + \ \gamma$ \
$ \& $ \  $ 0^{-+} \rightarrow h_{1} \ + \ \gamma$). This would 
effect the photon signal especially in the mixed mode which has been
a topic of much current work and has been reviewed recently [6]. 
This shall require redoing the calculations
for the photon signals [6]. In fact the time delay in the photon 
emission (due to finite life time of the hybrid) may perhaps
be accessible through a proper study within the
field of photon interferometry [6].

In summary, we show that in QGP at finite temperatures
 the dynamics of the system 
prefers to generate hybrids during hadronisation as opposed
to mesons. We suggest that this then constitutes a novel
signal of QGP. 
One also finds that the strong decay to mesons is forbidden
and hence these hybrids will decay electromagnetically, significantly
modifying the photonic signals of QGP.

\vskip 4.0 cm

We would like to thank Dr. M. G. Mustafa for giving 
us the computer code used herein and helping us in computation in 
the initail stages.

\vfill

\newpage

\noindent {\centerline {\bf CAPTIONS}}
\noindent {\bf Table 1} \\
The equillibrium radius$(R)$ 
 and the energy $(E)$ of the hybrid and pure mesonic
system is given for bag constant $ B^{1/4} = 250 $ MeV.

\vskip 7.0 cm

\noindent {\bf Figure 1.} \\
The variation of the free energy of the hybrid
and the pure mesonic systems with radius $R$ are displayed
at a particular temperature $ T = 183 $ MeV and
 $ B^{1/4} = 250 $ MeV.

\vfill

\newpage

\begin{table}
\centerline {\bf Table 1} 
\vskip 0.2 in
\begin{center}
\begin{tabular}{|c|c|c|c|}
\hline
System & T(MeV) & R(fm) & E(GeV) \\
\hline
($q\bar{q}g$) & 183 & 0.807 & 4.476 \\
\hline
($q\bar{q}$) & 183 & 0.658 & 2.426   \\ 
\hline
\end{tabular}
\end{center}
\end{table}
\end{document}